# Strain-driven light polarization switching in deep ultraviolet nitride emitters

T. K. Sharma,[1,2] Doron Naveh, and E. Towe[3]

Department of Electrical and Computer Engineering, Carnegie Mellon University
5000 Forbes Avenue, Pittsburgh, PA-15213, USA

Residual strain plays a critical role in determining the crystalline quality of nitride epitaxial layers and in modifying their band structure; this often leads to several interesting physical phenomena. It is found, for example, that compressive strain in $Al_xGa_{1-x}N$ layers grown on $Al_yGa_{1-y}N$ ($x<y$) templates results in an anti-crossing of the valence bands at considerably much higher Al composition than expected. This happens even in the presence of large and negative crystal field splitting energy for $Al_xGa_{1-x}N$ layers. A judicious magnitude of the compressive strain can support vertical light emission (out of the c-plane) from $Al_xGa_{1-x}N$ quantum wells up to $x \approx 0.80$, which is desirable for the development of deep ultraviolet light-emitting diodes designed to operate below 250nm with transverse electric polarization characteristics.

**PACS**: 78.66.Fd  78.66.-W  78.66.Bz  78.67.De  68.60.Bs

Deep ultraviolet (UV) nitride emitters have recently attracted a great deal of attention because of the possibility of developing more efficient, portable, and safer light sources as alternatives to conventional excimer and mercury lamps.[1-6] However, a major obstacle has been the extremely low external quantum efficiency (EQE) of nitride emitters. A significant drop in EQE is usually observed as emission wavelengths approach the deep UV regime.[5] The EQE for light-emitting diodes (LEDs) is defined as a product of the internal quantum efficiency (IQE), carrier injection efficiency (CIE) and the light extraction efficiency (LEE).[1,3-5] The internal quantum efficiency of a light-emitter is related to the crystalline quality of epitaxial layers where a large number of defects and dislocations leads to inferior device characteristics.[5,7] Inadequate carrier concentration in p-type AlGaN barrier layers and significantly reduced band discontinuities at key heterojunction interfaces are considered to be the major issues that affect carrier injection efficiency.[3-5,7] There are ongoing efforts to improve both the IQE and the CIE by using different methods. For example, an output power of 5(100) mw has been demonstrated under electrical(electron beam) injection for devices operating at ~250nm, respectively.[1,3] Similarly, low values of LEE (lower than 8%) are reported due to absorption of the UV light at the p-side electrode.[3] Furthermore, it is now well understood that light emission from Al-rich AlGaN layers grown on c-plane Sapphire is predominantly polarized along the c-axis; this makes light extraction from LEDs a formidable task.[8-10] Several strategies like interconnected microdisk LED architectures and two-dimensional photonic crystal designs have been suggested to enhance the light extraction efficiency of deep UV LEDs.[11,12]

Nevertheless, there is a fundamental question that has been intriguing researchers in the nitride community. It is found that light emitted by AlGaN epitaxial layers switches its polarization characteristics from transverse electric (TE) to transverse magnetic (TM) mode at some critical Aluminum composition. For example, Nam et al.[8] reported that the emitted light switches its polarization from TE to TM mode for $x > 0.25$ in $Al_xGa_{1-x}N$ epilayers grown on c-plane Sapphire. Similarly, Hazu et al.[13] reported polarization switching for Al content between $x=0.25$ and $x=0.32$ in $Al_xGa_{1-x}N$ epilayers on m-GaN substrates. In yet another paper, Kawanishi et al.[10] reported polarization switching at $x \approx 0.36$-0.41 in $Al_xGa_{1-x}N$ quantum wells on AlN/SiC templates. On the other hand, Ikeda et al.[14] reported polarization switching in $Al_xGa_{1-x}N$ epilayers on GaN/Sapphire templates, however at a much lower value of $x=0.125$. They argued that although the polarization switching should have been observed at $x=0.032$ for unstrained $Al_xGa_{1-x}N$ epilayers but it was observed at higher Al-content due to finite residual strain in the AlGaN epilayers. Recently, a similar argument was invoked by Banal et al.[2] to explain the polarization properties of light emitted by $Al_xGa_{1-x}N$/AlN quantum well (QW) structures grown on AlN/Sapphire templates; in this case the polarization switching was observed at $x \sim 0.80$. Using a qualitative model, they were able to explain some of their data. However, an ambiguity still exists on the critical value of Al composition in the $Al_xGa_{1-x}N$ epilayers where one expects to observe the polarization switching behavior. More recently, Hirayama et al.[3] have reported that they are able to observe vertical emission of light (out of the c-plane) up to x=0.83 for $Al_xGa_{1-x}N$ LEDs. This result appears to indicate that the polarization switching behavior may not be that important in deep UV AlGaN emitters. At the moment, there is neither clarity nor consensus on the subject. However, a lot of scattered data is available in the literature. These data appear to indicate that the critical Al composition at which the emitted light switches its polarization characteristics varies from x=0.125 to x=0.83. In this article, we discuss the fundamental physics related to the anti-crossing of valence bands and to the polarization switching of light emitted from $Al_xGa_{1-x}N$ epilayers. A quantitative analysis of the experimental data is essential for gaining a clear understanding of the switching of polarization characteristics of light emitted from $Al_xGa_{1-x}N$ epilayers.

---

[1]Corresponding Author: E-mail: tksharma@cmu.edu, Tel: +1-412-268-6726
[2]On leave from Semiconductor Laser Section, Raja Ramanna Centre for Advanced Technology, Indore -452 013 (M.P), India. Email: tarun@rrcat.gov.in
[3]Email:towe@cmu.edu

In our analysis, the modification of the electronic band structure and hole effective mass due to strain was taken into account using a 6x6 **k.p** formalism based on the Bir-Pikus Hamiltonian. This approach is now widely accepted in the nitride community.[13,15-19] All the materials parameters used in our calculations were taken from the literature.[19,20] The required parameters for the ternary alloys were interpolated by using appropriate values of the relevant binary GaN and AlN nitride materials. We then carried out meticulous quantum mechanical calculations aimed at estimating the ground state emission wavelength of wurtzite $Al_xGa_{1-x}N/Al_zGa_{1-z}N$ (x < z) QW structures suitable for the active regions of deep UV emitters. We numerically solve the Schrödinger equation for finite, square potential wells in the conduction and valence bands using the envelope function approximation. At the Brillouin zone center (Γ-point), the conduction band (CB) of AlGaN is composed of atomic s-orbitals with wave functions of $|S\rangle$ symmetry, while the three uppermost valence bands (VBs) are made up of p-orbitals with the wave functions being a combination of $|X\rangle$, $|Y\rangle$ and $|Z\rangle$ symmetry. A downward transition involving electrons in an $|S\rangle$-like CB state and holes in an $|X\rangle$-like, $|Y\rangle$-like, or $|Z\rangle$-like VB state would emit light polarized along the x-, y-, or z Cartesian directions, respectively. Here, the c-axis of the AlGaN wurtzite crystal is assumed to be parallel to the z-axis. We adopt a nomenclature wherein the strain-modified excitonic transitions are labeled **T₁**, **T₂**, and **T₃** in order of increasing energy. Their polarization properties are determined by the relative oscillator strength components $f_{ij}$ with i =1, 2, 3 and j = x, y, z representing the three transitions and their polarization components, respectively. The values of the matrix elements $|\langle S|p_x|X\rangle|^2$, $|\langle S|p_y|Y\rangle|^2$, and $|\langle S|p_z|Z\rangle|^2$ in AlGaN are predicted to be nearly equal, leading to the following oscillator strength sum rules $\sum_{i=1}^{3} f_{i,j} = 1$ and $\sum_{j=x}^{z} f_{i,j} = 1$.[16,19]

Because of the negative value of the crystal field splitting in AlN, the valence bands are expected to switch their character at some critical Al composition in the $Al_xGa_{1-x}N$ layers.[2,14] However, nitride epilayers grown on foreign substrates like Sapphire or SiC are prone to posses some residual strain.[19,20] The critical Al composition in the $Al_xGa_{1-x}N$ layers therefore depends on the magnitude of the residual strain as indicated by Ikida et al.[14] Using a qualitative model for the strain dependence of the crystal field splitting, Banal et al.[2] predicted that the bands should switch character at x=0.60 for $Al_xGa_{1-x}N$ layers grown on unstrained AlN. However, they observed switching of the polarization characteristics at x ~0.80, which was explained by assuming the presence of residual compressive strain in the AlN templates. If this is true, then $Al_xGa_{1-x}N$ (x>0.60) QWs grown on strain-free AlN templates/substrates might not be suitable for developing efficient deep UV LEDs with TE-mode characteristics. However, this contradicts recent observations by Hirayama et al.[3] who reported vertical emission of light (out of the c-plane) for aluminum compositions up to x=0.83. It should be noted that LED structures in the deep UV range are not grown on GaN or AlN, but on Al-rich $Al_xGa_{1-x}N$ templates/buffer layers to accommodate strain-free thick barrier layers.[3-5,7] It is therefore important to raise the issue of what the critical aluminum composition is at which the two valence bands switch their character for $Al_xGa_{1-x}N$ layers grown on $Al_yGa_{1-y}N$ template (x<y). With this in mind, we have carried out detailed theoretical calculations as described earlier and a typical example of the results is shown in Fig. 1 (a). Here, we plot the band gaps **E₁**, **E₂** and **E₃** at the Γ-point corresponding to the three valence bands **VB₁**, **VB₂** and **VB₃** of an $Al_xGa_{1-x}N$ ternary alloy grown on $Al_{0.50}Ga_{0.50}N$ free-standing substrate/template as a function of aluminum mole fraction, x, or equivalently strain. One notes that the two uppermost valence bands remain more or less degenerate (corresponding to **E₂-E₁**<15meV) for x values less than ~0.40; whereas for x >0.40; the degeneracy is lifted, and the band gap **E₃** becomes the smallest one. Well below the lattice-matching composition, there is an anti-crossing of the bands, even for compressive strain in the $Al_xGa_{1-x}N$ alloy; the two valence bands corresponding to $E_2$ and $E_3$, essentially switch their character.

Another important consequence of the strain-induced band structure modifications in Al-rich AlGaN QWs is a strong preference for TM polarization of the emitted light. We summarize the essential features of the impact on polarization in Fig. 1(b), where we plot the relative oscillator strength of the lowest excitonic transition ($T_1$ or $T_3$ depending on which band gap, $E_1$ or $E_3$, is lower) under strain for light polarized along the x-, y-, and z-direction for the $Al_xGa_{1-x}N$ layer. For purposes of comparison, we also show the additive values of the relative oscillator strength for the two components of polarization; in this case (X+Y), which determine the maximum fraction of light traveling perpendicular to the c-plane, for example in LEDs grown on c-plane Sapphire. For an aluminum composition of x > 0.40 in an $Al_xGa_{1-x}N$ QW, one notes that a good fraction of the emitted light is polarized along the c-axis. This is highly inconvenient for light extraction from LEDs whose output is expected along the c-axis. It is a fact that high Al-content in the active regions of LEDs is desirable for light emission at short wavelengths. However, inappropriate structural designs can lead to highly inefficient devices due to strain-induced polarization switching effects. These effects are often not properly accounted for in the design of LED devices. This situation is further complicated by the presence of thick barrier or buffer layer whose lattice constant is vastly different than that of the template/substrate. The barrier or buffer layers are bound to have some residual strain (even for fully relaxed ones in some cases); the residual strain alters the band structure. In view of this, we plot the numerically calculated polarization switching characteristics of $Al_xGa_{1-x}N$ layers that are assumed to be grown on $Al_yGa_{1-y}N$ templates (for x < y) in Fig. 2. The interpolated values of the crystal field splitting energy of the $Al_xGa_{1-x}N$ alloy are also plotted on the same graph (right axis). For x ≈0.04, the crystal field splitting changes its sign, indicating a possible switching of polarization characteristics. However, the critical Al composition increases due to residual strain in the $Al_xGa_{1-x}N$ layers as indicated in Fig. 2. The polarization characteristics are expected to switch at x≈0.80 for $Al_xGa_{1-x}N$ layers grown on free-standing AlN templates/substrates. This supports the observations of Banal et al.,[2] who reported polarization switching at x≈0.80 for $Al_xGa_{1-x}N$/AlN QW structures. In Fig. 3, we plot the critical Al composition in $Al_xGa_{1-x}N$ layers that leads to polarization switching in such layers if these were grown on free-standing $Al_yGa_{1-y}N$ templates/substrates of certain aluminum composition, y. The curve for 50% (1%) oscillator strength (Osc) represents the critical Al composition to ensure that 50% (1%) of the generated light travels parallel to the c-axis. The curve for Osc=1% indicates that the emitted light is strongly TM-polarized. These calculations clearly show that the critical Al composition increases linearly with Al content in the $Al_yGa_{1-y}N$ templates/substrates. For comparison purposes, we have also plotted the calculated values of the critical Al composition obtained using the qualitative model from Ref. 2. The qualitative model appears to be reasonable for Al-rich $Al_yGa_{1-y}N$ templates (y>0.90); however, it overestimates the critical Al composition for $Al_xGa_{1-x}N$ layers grown on $Al_yGa_{1-y}N$ templates with y<0.90.

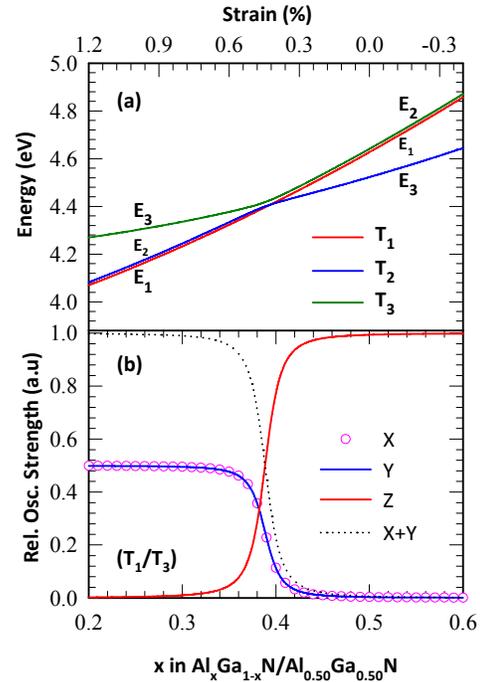

**Fig.1** (a) Excitonic band gaps $E_1$, $E_2$ and $E_3$ corresponding to three valence bands $VB_1$, $VB_2$ and $VB_3$ and (b) relative oscillator strength of the lowest excitonic transition ($T_1/T_3$) under strain for light polarized along the x-, y-, and z-directions plotted as functions of composition (bottom axis)/ residual strain (top axis), respectively for $Al_xGa_{1-x}N$ layers grown on $Al_{0.50}Ga_{0.50}N$ templates. Note the anti-crossing of the band gaps at **x~0.4**, corresponding to the strain-induced switching of the character of the valence bands.

It is interesting to compare and discuss the observations of Banal et al.[2] and Hirayama et al.[3], which we present in Table 1. According to the experimental results of Banal *et al.*[2], the critical Al composition for $Al_xGa_{1-x}N$ QW structures grown on AlN templates is ≈0.80; this value is in good agreement with our theoretical prediction as discussed here. Using a qualitative model, Banal et al.[2] predict that the polarization should switch from TM to TE for an $Al_xGa_{1-x}N$ QW structure with x=0.82 because of quantum confinement effects if the QW thickness is kept below 3nm. Using data from our model, we plot the energy values of ground state transitions corresponding to the three valence bands of **$Al_{0.82}Ga_{0.18}N/AlN$** QW structure in Fig. 4. The calculations indicate that the emitted light would be TM-polarized (where the **$T_3$** transition is the lowest one) for QW thicknesses larger than 5nm. However, the polarization should switch to TE-mode (when the **$T_1$** transition becomes the lowest one) for QW thicknesses below 5nm. Contrary to the qualitative model,[2] we find that the band gap **$E_3$**, corresponding to the valence band **$VB_3$** never becomes the largest of the three because of quantum confinement effects.

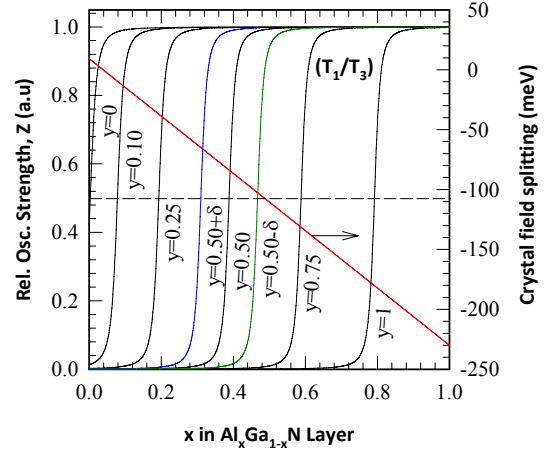

The experimental results of Hirayama et al.[3] shown in Table 1, are similarly in agreement with our calculations, except for the sample with x=0.83 where theory predicts TM-polarized light. For this particular QW sample, the valence bands switch their positions at a QW thickness of 2 nm because of quantum confinement effects. This leads to TE-polarized light. We note that the QW thicknesses used by Hirayama et al.,[3-5,7] are smaller than 2 nm, confirming our predictions. We point out that most deep UV LED structures are often grown on thick $Al_yGa_{1-y}N$ barrier/buffer layers which might be partially/fully relaxed. Any such relaxation processes would further influence the polarization characteristics of the emitted light. This is indicated in Fig. 2, where two additional curves for y=0.5 (with y = 0.5±δ) are plotted in order to include the residual strain (δ) in the $Al_yGa_{1-y}N$ barrier/buffer layer. Therefore, if a thick relaxed $Al_yGa_{1-y}N$ buffer layer with y=0.84 is used in an LED structure, the emitted light from an $Al_xGa_{1-x}N$ QW with x=0.67 should be TM-polarized as observed by Kawanishi et al.[10] In contrast to this, the emission characteristics of LEDs made by Hirayama et al.,[3] turn out to be TM-polarized for two of their samples when the buffer layers are assumed to be relaxed. Even with quantum confinement effects, the polarization cannot be switched back to the TE-mode. Furthermore, these devices were reported to have very low EQE values. It is important to note that the angular dependence of the light from the LEDs reported by Hirayama et al.[3] only measures the light fraction that is accommodated in the escape cone, which is not necessarily a good measure for the polarization characteristics of light emitted by a QW structure.

**Fig.2** Relative oscillator strength of the lowest excitonic transition (**$T_1/T_3$**) under strain for light polarized along the z-direction plotted as functions of Al compositions of **$Al_xGa_{1-x}N$** layers grown on **$Al_yGa_{1-y}N$** templates. Switching of the light polarization from TE to TM-mode occurs whenever the oscillator strength increases above 0.5, as shown by a dashed line which defines the critical Al composition of **$Al_xGa_{1-x}N$** layers grown on **$Al_yGa_{1-y}N$** templates. Effect of residual strain (δ=0.4%) in **$Al_{0.50}Ga_{0.50}N$** templates on the corresponding critical Al composition of **$Al_xGa_{1-x}N$** layers is also shown. The interpolated values of the crystal field splitting of the **$Al_xGa_{1-x}N$** layers are shown in the same graph (right axis).

| References | Template/ Substrate | 'y' in $Al_yGa_{1-y}N$ barrier/buffer layer | 'x' in $Al_xGa_{1-x}N$ Quantum well | Polarization Experimental (TE/TM) | Polarization Theory (TE/TM) |
|---|---|---|---|---|---|
| **[2]** | AlN/Sapphire | 1 | 0.69 | TE | TE |
|  |  | 1 | 0.82 | TE/TM* | TE/TM* |
|  |  | 1 | 0.91 | TM | TM |
| **[3]** | AlN/Sapphire | 0.89 | 0.83 | TE | TE/TM* |
|  |  | 0.84 | 0.74 | TE | TE |
|  |  | 0.76 | 0.61 | TE | TE |

*Quantum confinement changes the polarization of emitted light from TM to TE mode.

**Table 1** Experimental and theoretical light polarization characteristics of **$Al_xGa_{1-x}N$** QW structures grown on **$Al_yGa_{1-y}N$** buffer/barrier layers on AlN/Sapphire templates. TE/TM stands for transverse electric/magnetic light polarization respectively.

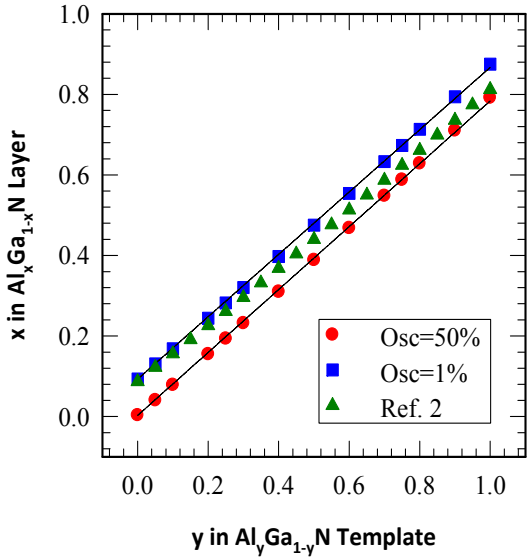

**Fig. 3** The critical Al composition in $Al_xGa_{1-x}N$ layers leading to a switching of the light polarization characteristics (Osc=50%) plotted as function of compositions of $Al_yGa_{1-y}N$ templates. For comparison, the values of critical Al compositions obtained from a qualitative model (Ref. 2). and the Al composition that supports strongly TM-polarized light (Osc=1%) emission is also shown in the same graph.

In the foregoing treatment and discussion, we have ignored the effects of polarization-induced electric fields for simplicity; these are known to lead to a red-shift in the emission wavelength and a reduction in device efficiency due to the separation of the electron and hole wave functions. In shallow wells, the polarization-induced electric field can facilitate thermal ejection of charge carriers from the QWs. However, these effects can be reasonably suppressed under heavy current injection conditions for very thin QWs (< 2 nm) such as those that are common in deep UV nitride emitters. Furthermore, it has already been predicted that the critical Al-composition remains largely unaffected by the electric field for thin QWs.[2] In any event, these effects are known to be less prominent for AlGaN QWs emitting in the UV range compared to the InGaN QWs emitting in the blue-green spectral band.[21-23]

Finally, we plot the calculated values of the relative oscillator strength for the lowest excitonic transition ($T_1/T_3$) for $Al_xGa_{1-x}N/Al_zGa_{1-z}N$ QWs grown on $Al_yGa_{1-y}N$ (x<z<y) templates as functions of the emission wavelength in Fig. 5. The values of the Al compositions (x, y and z) for different QW structures were taken from those corresponding to LED structures in the literature (Expt-1: Ref. 3, Expt-2: Ref. 7, Expt-3: Ref. 24). Note that the calculated values of the relative oscillator strength for the lowest excitonic transition decrease rapidly as the wavelength of 200 nm is approached for Al-rich AlGaN QWs. For comparison, the experimentally measured values of the EQE, taken from the same references are also plotted on the same graph (right axis).[3,7,24] In addition, we have included the reported EQE values for AlInGaN and AlN LEDs as benchmarks. This figure illustrates the fact that reduction of the oscillator strength as the wavelength becomes shorter in $Al_xGa_{1-x}N$ QWs tracks the decrease in the EQE values. This correlation reasonably explains the drop in EQE values that is usually observed in the deep UV LEDs. It is related to the dominance of TM-polarized light that is difficult to extract from LEDs grown on c-plane templates/substrates. This issue has not been properly addressed in deep UV LEDs where the precipitous decrease in EQE is usually associated with problems in epitaxial growth, such as difficulties in achieving high electrical conductivity in p-AlGaN barrier layers.[3-5,7] Theoretical calculations predict that the EQE of AlN homojunction LEDs should be lower by about three orders of magnitude compared to the AlGaN LEDs operating at 280 nm, even when problems in epitaxial growth are resolved. In order to develop efficient deep UV LEDs, it is therefore essential to design the QW structures with compressive residual strain by choosing suitable $Al_yGa_{1-y}N$ templates/buffers that ensure TE-polarized light emission. One optimal LED design for TE-polarized light emission at ~240 nm, for example, might include an $Al_{0.62}Ga_{0.38}N/Al_{0.80}Ga_{0.20}N$ QW structure in the active region grown on $Al_{0.85}Ga_{0.15}N$ template/substrate. For TE-mode emission at the shorter wavelength of 220 nm, the QW could be composed of an $Al_{0.75}Ga_{0.25}N/Al_{0.90}Ga_{0.10}N$ structure grown on an AlN template/substrate. The metrics of these two optimized polarization properties (OPP) LEDs are indicated (with stars) in Fig. 5.

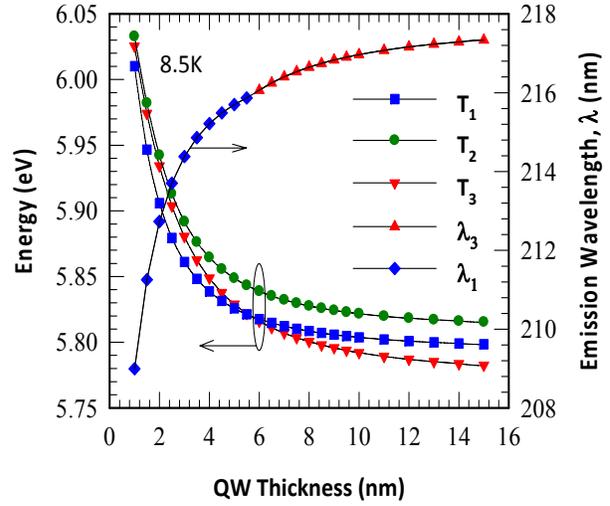

**Fig.4** Ground state energies of $T_1$, $T_2$, and $T_3$ excitonic transitions of an $Al_{0.82}Ga_{0.18}N/AlN$ QW structure corresponding to the three band gaps $E_1$, $E_2$, and $E_3$ plotted as a function of QW thickness. The emission wavelength corresponding to the lowest excitonic transition is also indicated on the same graph (right axis). TM and TE polarization of emitted light is indicated by $\lambda_3$ and $\lambda_1$ plots corresponding to $T_3$ and $T_1$ excitonic transitions respectively, note that their positions switch at a QW thickness of 5 nm.

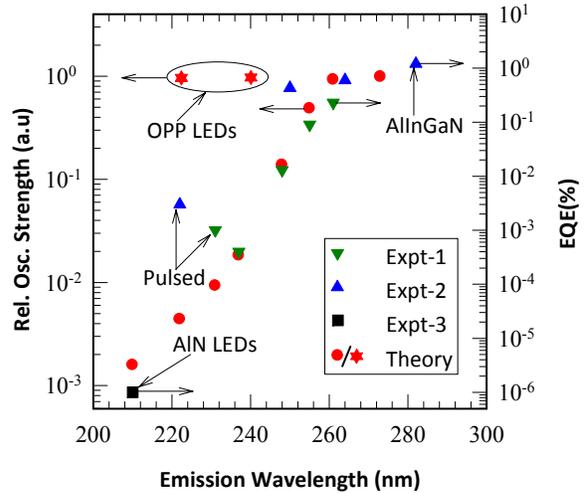

**Fig.5** Relative oscillator strength for the lowest excitonic transition ($T_1/T_3$) for AlGaN QWs plotted as a function of emission wavelength. For comparison, values of the external quantum efficiencies (EQE) are also plotted on the same graph (Expt-1:Ref-3, Expt-2:Ref-7, Expt-3:Ref-24). The relative oscillator strengths for two LED structures whose designs are optimized for TE-polarization are also shown (OPP LEDs).

In summary, we find that there is a linear relationship between the critical Al compositions in **$Al_xGa_{1-x}N$** epilayers used in the active region of LEDs and the Al fraction of respective **$Al_yGa_{1-y}N$** templates/substrates. The polarization characteristics of light emitted from such active layers switch from TE to TM-mode at the critical Al compositions. Furthermore, the critical Al composition of the active layer has a dependence on residual strain in the **$Al_yGa_{1-y}N$** templates. An accurate quantitative estimate of the strain-driven critical Al composition in **$Al_xGa_{1-x}N$** layers is thus desired where even quantum confinement effects might play a crucial role in deciding the emission characteristics of certain QW structures. Since strain-induced polarization switching of emitted light can lead to a reduction in the light extraction efficiency of LEDs, it is imperative to determine *a priori* what the optimal device design parameters should be. We suggest two QW structures that are designed to emit TE-polarized in the deep UV, with a possibility of achieving EQE values similar to those obtained in 280 nm LEDs. We note that while higher Al-content in QWs might lead to even shorter emission wavelengths (below 220 nm), this comes at a price that includes poor carrier confinement, lower extraction of the TM-polarized light, and higher threshold current densities. We also note that the importance of incorporating a reasonable number of ionized acceptor impurities in Al-rich AlGaN barrier layers is of paramount importance. However, the issue of the critical Al composition for the polarization switching is of a fundamental nature, and must be taken care of in the design of deep UV LEDs.

The authors acknowledge the financial support of the U.S. Defense Advanced Research Projects Agency through the U.S. Air Force at Hanscom Air Force Base in Bedford, MA for the work on nitride semiconductors (Grant No. FA-871808-c-0035). TKS thank Dr. Sandip Ghosh for useful discussions.

[1] T. Oto, R. G. Banal, K. Kotaoka, M. Funato and Y. Kawakami, Nature Photon. 26 September, 2010. doi:10.1038/nphoton.2010.220.
[2] R. G. Banal, M. Funato and Y. Kawakami, Phy. Rev. B **79**, 121308R (2009).
[3] H. Hirayama, N. Noguchi and N. Kamata, Appl. Phys. Express **3**, 032102 (2010).
[4] H. Hirayama, Y. Tsukada, T. Maeda and N. Kamata, Appl. Phys. Express **3**, 031002 (2010).
[5] H. Hirayama, T. Yatabe, N. Noguchi, T. Ohashi and N. Kamata, Electronics and Communications in Japan **93**, 748 (2010).
[6] W. Sun, M. Shatalov, J. Deng, X. Hu, J. Yang, A. Lunev, Y. Bilenko, M. Shur and R. Gaska, Appl. Phys. Lett. **96**, 061102 (2010).
[7] H. Hirayama, S. Fujikawa, N. Noguchi, J. Norimatsu, T. Ta-kano, K. Tsubaki, and N. Kamata, Phys. Status Solidi A **206**, 1176 (2009).
[8] K. B. Nam, J. Li, M. L. Nakarmi, J. Y. Lin, and H. X. Jianga, Appl. Phys. Lett. **84**, 5264 (2004).
[9] H. Kawanishi, M. Senuma, M. Yamamoto, E. Niikura and T. Nukui, Appl. Phys. Lett. **89**, 081121 (2006).
[10] H. Kawanishi, M. Senuma, and T. Nukui, Appl. Phys. Lett. **89**, 041126 (2006).
[11] J. Shakya, K. H. Kim, J. Y. Lin, and H. X. Jiang, Appl. Phys. Lett. **85**, 142 (2004).
[12] J. Shakya, K. Onabe, K. H. Kim, J. Li, J. Y. Lin, and H. X. Jiang, Appl. Phys. Lett. **86**, 091107 (2005).
[13] K. Hazu, T. Hoshi, M. Kagaya, T. Onuma and S. F. Chichibu, J. Appl. Phys. **107**, 033701 (2010).
[14] H. Ikeda, T. Okamura, K. Matsukawa, T. Sota, M. Sugawara, T. Hoshi, P. Cantu, R. Sharma, J. F. Kaeding, S. Keller, U. K. Mishra, K. Kosaka, K. Asai, S. Sumiya, T. Shibata, M. Tanaka, J. S. Speck, S. P. DenBaars, S. Nakamura, T. Koyama, T. Onuma and S. F. Chichibu, J. Appl. Phys. **103**, 089901 (2008).
[15] M. Suzuki, T. Uenoyama, and A. Yanase, Phy. Rev. B **52**, 8132 (1995).

[16] M. Suzuki and T. Uenoyama, Jpn. J. Appl. Phys. **35**, 543 (1996).
[17] A. Shikanai, T. Azuhata, T. Sota, S. Chichibu, A. Kuramata, K. Horino and S. Nakamura, J. Appl. Phys. **81**, 417 (1997).
[18] S. Ghosh, P. Waltereit, O. Brandt, H. T. Grahn, and K. H. Ploog, Phy. Rev. B 65, 075202 (2002).
[19] J. Bhattacharyya, S. Ghosh, and H. T. Grahn, Phys. Status Solidi B **246**, 1184 (2009).
[20] I. Vurgaftman and J. R. Meyer, J. Appl. Phys. **94**, 3675 (2003).
[21] H. Yoshida, Y. Yamashita. M. Kuwabara and H. Kan, Nature Photon. **2**, 551 (2008).
[22] D. F. Feezell, M. C. Schmidt, S. P. DenBarrs, and S. Nakamura, MRS Bull. **34**, 318 (2009).
[23] K. Kazlauskas, G. Tamulaitis, J. Mickevièius, E. Kuokštis, A. Žukauskas, Y. C. Cheng, H. C. Wang, C. F. Huang, and C. C. Yang, J. Appl. Phys. **97**, 013525 (2005).
[24] Y. Taniyasu, Makoto Kasu, and T. Makimoto, Nature **441**, 325 (2006).